\documentclass[twocolumn]{emulateapj}
\submitted{Science with Parkes @ 50 Years Young, 31 Oct. -- 4 Nov., 2011}
\shorttitle{Circumstellar masers in the Magellanic Clouds}
\shortauthors{Jacco Th. van Loon}

\newcommand\actaa{Acta Astron.}%

\begin{document}
\title{Circumstellar masers in the Magellanic Clouds}
\author{Jacco Th. van Loon}
\affil{Lennard-Jones Laboratories, Keele University, ST5 5BG, UK}
\email{j.t.van.loon@keele.ac.uk}
\begin{abstract}
The nearby dwarf irregular galaxies the Large and Small Magellanic Clouds have
metallicities of about half and a fifth solar, respectively, which offers the
unique opportunity to study astrophysical processes as a function of
metallicity. Masers in the outflows from evolved stars allow to measure the
wind speed, vital to derive mass-loss rates and test wind driving mechanisms.
The metallicity dependence of the wind speed in particular allows us to make
inferences about dust formation and mass loss in the early Universe. I will
review past surveys for circumstellar OH, water, and SiO masers in the
Magellanic Clouds (and provide a literature review of interstellar masers). I
will then discuss the way these measurements have influenced our understanding
of mass loss, and end with outlining the prospects for future surveys for OH
masers in the Magellanic Clouds.
\end{abstract}
\keywords{masers
     --- stars: AGB and post-AGB
     --- stars: mass-loss
     --- supergiants
     --- stars: winds, outflows
     --- Magellanic Clouds}
\section{The astrophysical significance of masers in the Magellanic Clouds}

Microwave amplified stimulated emission radiation (maser) arises when emission
from a transition within a molecule is amplified when it causes subsequent
stimulated emission from the same transition within another such molecule. For
this to happen, a population inversion needs to be instated, so the
meta-stable higher energy level of the masing transition becomes populated
preferentially over the energy level to which it decays. The mechanism
responsible for this ``pumping'' can be either through the absorption of
radiation in transitions to levels from which the upper level of the masing
transition can be populated, or through collisions with other particles.
Masers therefore require specific conditions to be met; conversely, the
radiation is highly an-isotropic.

The result is extremely bright emission linked to specific regions and
kinematics in the object (Elitzur, Goldreich, \& Scoville 1976; Chapman \&
Cohen 1986). In the outflows of cool evolved star winds -- Asymptotic Giant
Branch (AGB) stars or red supergiants -- prototypical hydroxyl (OH) masers
display a double-horned profile separated by twice the wind speed; water
(H$_2$O) masers probe accelerating parts of the wind near the dust formation
z\^one; while silicon-monoxide (SiO) masers tend to peak near the stellar
systemic velocity. The detection of masers in these objects thus allows the
measurement of the gas kinematics in the outflow, which informs us about its
driving mechanism -- believed to be due to radiation pressure on dust grains
-- and about the grain properties. In star forming regions, methanol
(CH$_3$OH) masers play a valuable r\^ole in locating warm cores.

The Magellanic Clouds offer us unique laboratories for studying astrophysical
processes due to their proximity (about 50 and 60 kpc for the Large and Small
Magellanic Clouds, respectively), different metal content (half and a fifth
solar, for respectively the LMC and SMC), and decent size and on-going star
formation resulting in sufficient examples of the intrinsically rare stages
during which masers are observed.

\section{The known circumstellar masers in the Magellanic Clouds}

\begin{table*}
\begin{center}
\caption{Literature on circumstellar masers in the Magellanic Clouds.}
\begin{tabular}{lllll}
\tableline\tableline
year  & authors         & telescope         & molecule & remarks         \\
\tableline
1986  & Wood et al.     & Parkes            & OH       & first detection \\
1992  & Wood et al.     & ATCA+Parkes       & OH       &                 \\
1996  & van Loon et al. & SEST              & SiO      & first detection \\
1998a & van Loon et al. & ATCA              & OH       &                 \\
1998b & van Loon et al. & Parkes            & H$_2$O   & first detection \\
2001b & van Loon et al. & Mopra+Parkes+SEST & H$_2$O + SiO &             \\
2004  & Marshall et al. & Parkes            & OH       &                 \\
\tableline
\end{tabular}
\end{center}
\end{table*}

I have listed in Table 1 an overview of the literature on circumstellar masers
in the Magellanic Clouds -- i.e.\ masers in the outflows from cool evolved
stars. Not only have there been few publications, the number of detected
masers (10) is so small that I can list them all (Table 2, along with a few
candidates based on their similar spectral types, pulsation periods and dust
envelopes). To date, no circumstellar masers are known in the SMC. For the
sake of this review I consider all other masers to be of ``interstellar''
origin, and only provide a list of references in relation to their detection
in the Magellanic Clouds (Table 3).

\begin{table*}
\begin{center}
\caption{Circumstellar masers in the Magellanic Clouds including candidate
OH/IR stars, with spectral type, pulsation period and OH peak intensity
(r.m.s.\ noise in parentheses).}
\begin{tabular}{lllrcl}
\tableline\tableline
IRAS         & other name  & SpT      & P (d) & OH peak (mJy) & remarks \\
\tableline
\multicolumn{6}{l}{\it Small Magellanic Cloud} \\
00483$-$7347 &               & late-M & 1859  &         (9)   &         \\
00486$-$7308 & GM\,103       & M4     & 1062  &          ?    &         \\
00591$-$7307 & HV\,11417     & M5e\,I & 1092  &         (8)   &         \\
\multicolumn{6}{l}{\it Large Magellanic Cloud} \\
04407$-$7000 &               & M7.5   & 1199  &         50    &         \\
04498$-$6842 &               & M10    & 1292  &         23    &         \\
04509$-$6922 &               & M10    & 1292  &        (17)   &         \\
04516$-$6902 &               & M9     & 1091  &        (11)   &         \\
04545$-$7000 &               &        & 1216  & \llap{1}40    &         \\
04553$-$6825 & WOH\,G064     & M7.5   &  841  & \llap{6}00    & H$_2$O, SiO \\
05003$-$6712 &               & M9     &  883  &         33    &         \\
05280$-$6910 & NGC\,1984-IR1 &        &       &         90    & H$_2$O  \\
05294$-$7104 &               & M8     & 1079  &        (13)   &         \\
05298$-$6957 & HS\,327E-IR1  &        & 1280  & \llap{2}40    &         \\
05329$-$6708 &               &        & 1262  & \llap{1}30    &         \\
05402$-$6956 &               &        & 1393  &         80    &         \\
05558$-$7000 &               &        & 1220  &         17    &         \\
\tableline
\end{tabular}
\end{center}
\end{table*}

\begin{table*}
\begin{center}
\caption{Literature on interstellar masers in the Magellanic Clouds.}
\begin{tabular}{lllll}
\tableline\tableline
year & authors               & telescope     & molecule  & remarks          \\
\tableline
1977 & Kaufmann et al.       & Itapetinga    & H$_2$O    & no detection     \\
1981 & Caswell \& Haynes     & Parkes        & OH        & first detection  \\
1981 & Haynes \& Caswell     & Parkes        & OH        &                  \\
1981 & Scalise \& Braz       & Itapetinga    & H$_2$O    & first detection  \\
1982 & Scalise \& Braz       & Itapetinga    & H$_2$O    &                  \\
1983 & Whiteoak et al.       & Parkes        & H$_2$O    &                  \\
1985 & Gardner \& Whiteoak   & Parkes        & OH        &                  \\
1986 & Whiteoak \& Gardner   & Parkes        & H$_2$O    &                  \\
1992 & Sinclair et al.       & Parkes        & CH$_3$OH  & first detection  \\
1994 & Ellingsen et al.      & ATCA+Parkes   & CH$_3$OH  &                  \\
1995 & Caswell               & ATCA+Parkes   & OH        & excited OH       \\
1996 & Beasley et al.        & ATCA          & CH$_3$OH  &                  \\
1997 & Brooks \& Whiteoak    & ATCA          & OH        &                  \\
2001 & van Loon \& Zijlstra  & Mopra         & H$_2$O    &                  \\
2002 & Lazendi\'c et al.     & ATCA+Canberra & H$_2$O    &                  \\
2004 & Brogan et al.         & ATCA          & OH       & supernova remnant \\
2005 & Roberts \& Yusef-Zadeh & ATCA         & OH       & supernova remnant \\
2006 & Oliveira et al.       & Parkes        & H$_2$O    &                  \\
2008 & Green et al.          & ATCA+Parkes   & CH$_3$OH + OH & excited OH   \\
2010 & Ellingsen et al.      & ATCA+Parkes   & CH$_3$OH + H$_2$O &          \\
\tableline
\end{tabular}
\end{center}
\end{table*}

Few circumstellar masers have been detected due to their faintness as well as
difficulties with the target selection. Targets were selected predominantly
from the IRAS point source catalogue, but many of the targets selected for the
first study by Wood et al.\ (1992) later turned out to be carbon stars --
which do not display masers -- or not cool evolved stars at all, resulting in
their low success rate of 6 out of 54. The second search, with the ATCA (van
Loon et al.\ 1998a) targetted two objects, one of which was detected
(IRAS\,04407$-$7000) while the other was a carbon star after all. The third
systematic survey, presented in Marshall et al.\ (2004), drew targets from the
same pool of candidates but these had by then been much better characterised.
OH/IR stars invariably have oxygen-rich dusty envelopes but also late-M
spectral types and long pulsation periods reflecting their extremely evolved
status. This selection yielded a 50 per cent success rate for the LMC targets,
but it was evident that deeper searches would be needed to detect most of the
candidates, and that some of the best candidates may still have escaped
identification.

Of the detected maser sources (and some of the LMC candidates), pulsation
periods were determined by Wood et al.\ (1992), dust chemical types by Trams
et al.\ (1999), mass-loss rates by van Loon et al.\ (1999, 2005), and spectral
types by van Loon et al.\ (1998a, 2005); their maser properties were
summarised by Marshall et al.\ (2004). The SMC sources were characterised by
Groenewegen \& Blommaert (1998) with pulsation periods by Soszy\'nski et al.\
(2011). Some sources are located in a star cluster, so we know the birth mass
-- due to mass loss the present-day mass will have diminished, which is partly
reflected in the long pulsation period: IRAS\,05280$-$6910 is a red supergiant
evolved from a 19-M$_\odot$ main-sequence star (Wood et al.\ 1992; van Loon,
Marshall, \& Zijlstra 2005) while IRAS\,05298$-$6957 used to be a 4-M$_\odot$
main-sequence star (van Loon et al.\ 2001a).

The extremely luminous and late-type dusty red supergiant IRAS\,04553$-$6825
(WOH\,G064; cf.\ Elias, Frogel, \& Schwering 1986) is the only OH/IR star in
the Magellanic Clouds in which besides OH (Wood et al.\ 1986, 1992) -- both
the usual 1612 MHz satellite line as well as the 1665 and 1667 MHz main lines
-- also both H$_2$O (van Loon et al.\ 1998b) and SiO (van Loon et al.\ 1996)
masers have been detected. Detection of the SiO maser revealed that what had
originally been thought to be OH emission from the receding part of the
circumstellar envelope actually is substructure in the approaching part of the
envelope, the emission from the receding part was detected later and found to
be much fainter (Marshall et al.\ 2004). This added to suspicions based on the
spectral energy distribution -- bright infrared emission from dust but
relatively little obscuration at optical wavelengths (Roche, Aitken, \& Smith
1993; van Loon et al.\ 1999) -- that the circumstellar envelope must resemble
a torus or a shell with bipolar cavities, a conjecture which was beautifully
confirmed with interferometric measurements by Ohnaka et al.\ (2008).

\section{Testing radiation driven wind theory}

There is strong empirical evidence for the mass-loss rates of cool evolved
stars to depend very little on metal content but for the dust content of those
outflows to scale approximately in proportion to metal content (Elias,
Frogel, \& Humphreys 1985; van Loon 2000; van Loon et al.\ 2008). Given this,
simple prescriptions for radiation-driven dusty winds yield predictions for
how the wind speed should depend on metal content and on luminosity (Habing,
Tignon, \& Tielens 1994; Elitzur \& Ivezi\'c 2001; Marshall et al.\ 2004):
\begin{equation}
v\ \propto\ \psi^{1/2}\ L^{1/4},
\end{equation}
where the dust:gas mass ratio $\psi\simeq1:200$ at solar metallicity. This
prediction was verified by comparing the measured wind speed as a function of
luminosity in the LMC and in the Galactic Centre (Marshall et al.\ 2004).

There is a suggestion (Fig.\ 16 in Marshall et al.\ 2004) that the wind speed
might in fact show a steeper dependence on luminosity in the LMC sample; this
might be due to differences in the gravitational acceleration around the dust
formation z\^one between stars of different masses and sizes, or possibly the
dust fraction is higher for the more massive progenitor stars. The H$_2$O
maser profiles of IRAS\,04553$-$6825 and IRAS\,05280$-$6910 clearly reveal the
acceleration of the envelope, possibly at a slower rate than seen in Galactic
objects (van Loon et al.\ 2001b). This could result from a lower dust content,
with consequently a larger drift velocity between the dust grains and the gas
fluid.

A similar discrepancy in parameter dependence was found by Marshall et al.\
(2004) between the mass-loss rates determined from the dust emission and those
from the OH maser intensity following the recipe by Baud \& Habing (1983) as
presented by van der Veen \& Rugers (1989); this suggests that the OH maser
intensity may in fact measure the dust emission responsible for the maser
pumping, rather than the gas mass (Zijlstra et al.\ 1996).

\section{IRAS\,05298$-$6957: a case study}

IRAS\,05298$-$6957 is the second-strongest OH maser source in the LMC, with
the most ``classic'' maser profile, and also located in a star cluster. In
Fig.\ 1 I plot its spectral energy distribution, with near-infrared photometry
from van Loon et al.\ (2001a) and Wood et al.\ (1992), and 24- and 70-$\mu$m
measurements obtained with the {\it Spitzer} Space Telescope (cf.\ van Loon et
al.\ 2010). Overplotted is a 3000\,K blackbody representing the central star
(its effective temperature is not known but it is likely a late-M type star),
and the spectral energy distribution resulting from radiation transfer through
a dusty circumstellar envelope. The latter was computed using the {\sc dusty}
code (Nenkova, Ivezi\'c, \& Elitzur 2000), assuming a standard size
distribution (Mathis, Nordsieck, \& Rumpl 1977) of silicate grains (Draine \&
Lee 1984) that attain a temperature of 1000 K at the inner edge of the dust
envelope. To match the measurements the computed model had to be scaled to a
luminosity of 40\,000 L$_\odot$ and, for a dust:gas mass ratio of 1:500, a
mass-loss rate of $2\times10^{-4}$ M$_\odot$ yr$^{-1}$.

\begin{figure}
\resizebox{\hsize}{!}{\includegraphics{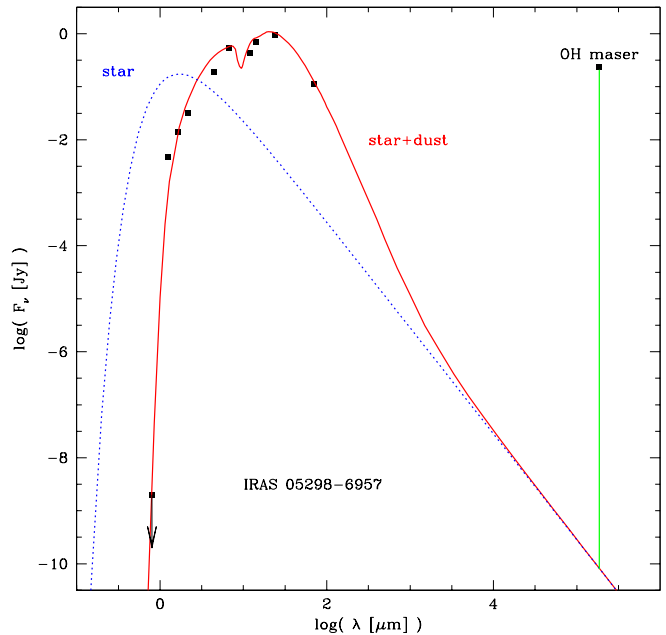}}
\caption{Spectral energy distribution of IRAS\,05298$-$6957 including the OH
maser.}
\end{figure}

\begin{figure*}
\centerline{\includegraphics[width=145mm]{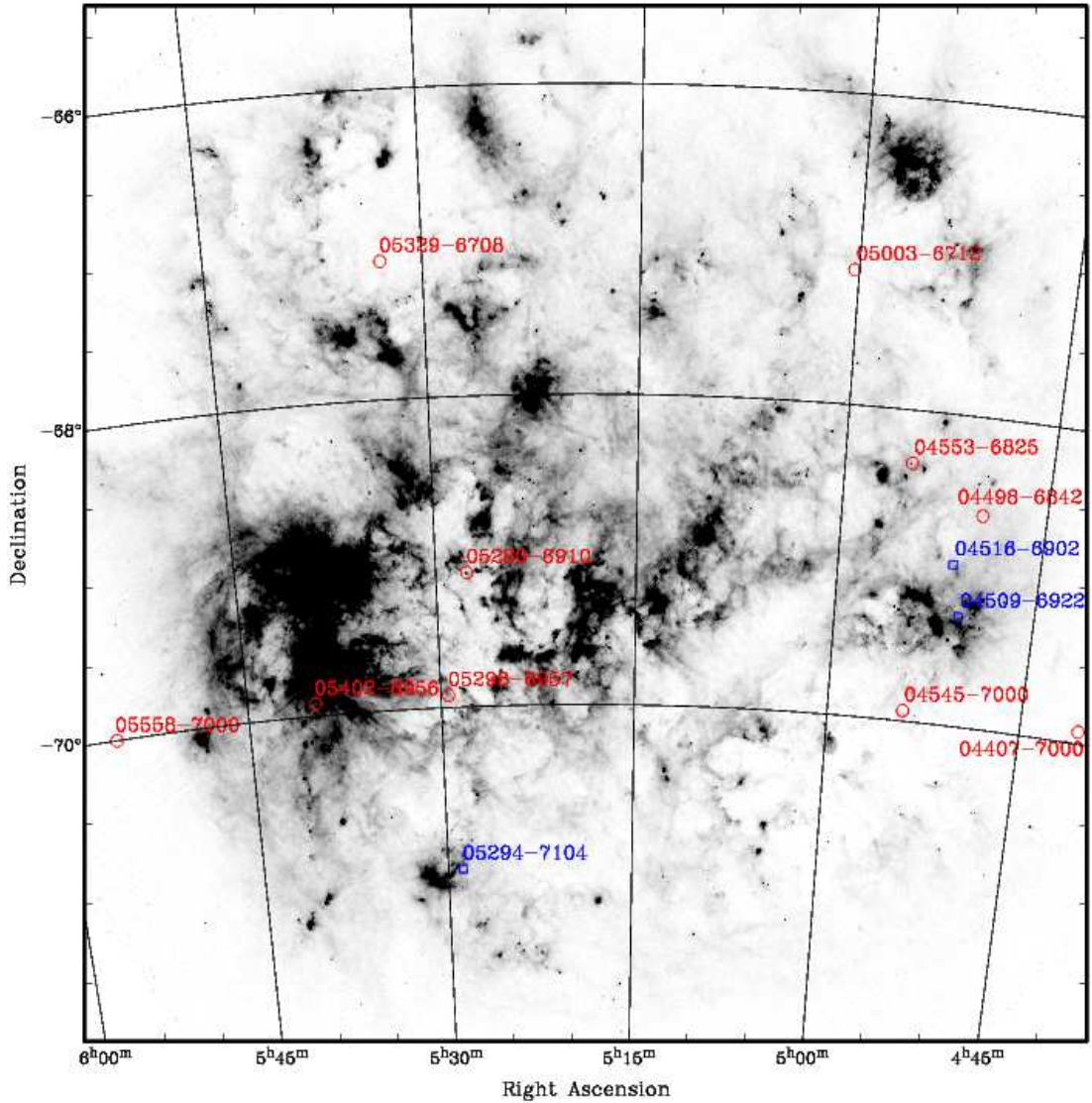}}
\caption{Detected circumstellar OH masers (red circles) and candidate OH/IR
stars (blue squares) in the Large Magellanic Cloud, on a 24-$\mu$m {\it
Spitzer} image \citep{Meixner06}.}
\end{figure*}

\begin{figure*}
\centerline{\includegraphics[width=135mm]{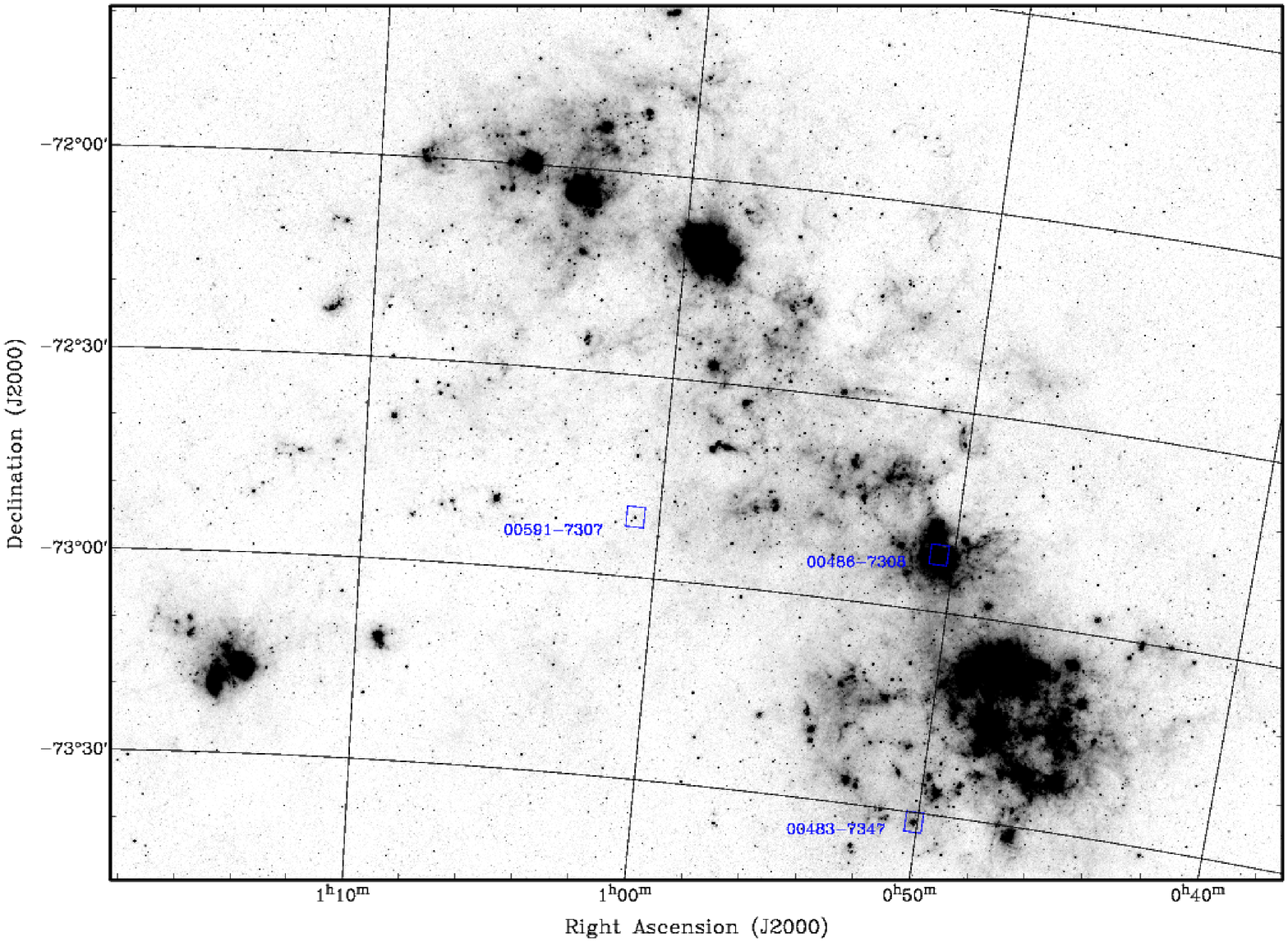}}
\caption{Candidate OH/IR stars in the Small Magellanic Cloud, on a 24-$\mu$m
{\it Spitzer} image \citep{Gordon11}.}
\end{figure*}

At radio wavelengths, the brightness is dominated by the stellar photosphere,
but the OH maser is more than a billion times ({\it sic} !) brighter, reaching
similar levels as the peak of the spectral energy distribution at mid-infrared
wavelengths (Fig.\ 1). The {\sc dusty} model predicts a wind speed of 7.2 km
s$^{-1}$, which is somewhat slower than the measured 10.5 km s$^{-1}$
(Marshall et al.\ 2004). Possibly the grain properties are a little different
or the dust:gas mass ratio is a little higher than assumed (raising it to
$\psi\simeq1:250$ would explain it). This exercise serves to demonstrate that
measurement and modelling of the wind speed hold great promise to further our
understanding of these dusty winds.

\section{Future prospects}

The current best candidate OH/IR stars in both Magellanic Clouds (Table 2)
very likely exhibit detectable OH emission not too far below the detection
limits of past searches. Efforts are underway with Parkes and the ATCA to
detect these. However, the IRAS survey was heavily confused in the Magellanic
Clouds. Figs.\ 2 and 3 show the locations of the detected and candidate OH/IR
stars on more recent {\it Spitzer} Space Telescope maps of the 24-$\mu$m
emission, the wavelength where the relatively warm circumstellar dust emission
peaks. Clearly, these sources are generally located in regions without strong
diffuse emission from small interstellar grains. Especially red supergiant
OH/IR stars would more often be found in dusty regions as they are not much
older than the typical lifetime of the molecular cloud complexes in which they
formed. Thus the {\it Spitzer} maps may harbour excellent targets, most of
which await identification and characterisation.

The situation will improve dramatically with the advent of Square Kilometre
Array (SKA) pathfinder experiments -- viz.\ the Australian SKA Pathfinder
(ASKAP) and Southern African MeerKAT arrays (van Loon, GASKAP Team \& MagiKAT
Team 2010). In particular the GASKAP survey combines high sensitivity (200-hr
integrations) with large instantaneous synthesized field (30 square degrees)
to reach the required survey speed and depth for a blind survey of the
Magellanic Clouds, thus avoiding the need for pre-selection of targets. The
synthesized beam of $<20^{\prime\prime}$ greatly reduces the level of beam
dilution and possible confusion at the low sensitivity that will be achieved
(r.m.s.\ noise well below a mJy). Both the 1612 and 1665/1667 MHz transitions
will be searched. Based on the detected masers in the LMC and Milky Way, and
employing scaling relations of between the OH maser intensity and mid-infrared
brightness, it is estimated that the GASKAP survey will uncover dozens of new
OH/IR stars in the LMC (besides obtaining better quality data on the already
known masers), and the first sample of OH/IR stars in the SMC. Unprecedented
tests of the driving of cool evolved star winds will become possible,
especially with regard to its dependence on metal content.


\end{document}